\documentclass[journal,10pt,doublecolumn,twoside, a4paper]{IEEEtran}
\usepackage{cite}
\usepackage{amsmath,amssymb,amsfonts,amsthm}
\usepackage{mathrsfs}
\usepackage{float}
\usepackage{hyperref}
\usepackage{graphicx}
\usepackage{textcomp}
\usepackage{xcolor}
\usepackage{xparse}
\usepackage{subfiles}
\usepackage{arydshln}
\usepackage{algorithm}
\usepackage{algpseudocode}

\newcommand{\R}{\mathbb{R}}

\newcommand{\bmat}[1]{\begin{bmatrix}#1\end{bmatrix}}

\newcommand{\ip}[2]{\left\langle #1, #2 \right\rangle}
\newcommand{\mcl}[1]{\mathcal{#1}}
\newcommand{\mbf}[1]{\mathbf{#1}}
\newcommand{\RL}[0]{\R L_2}

\def\BibTeX{{\rm B\kern-.05em{\sc i\kern-.025em b}\kern-.08em
    T\kern-.1667em\lower.7ex\hbox{E}\kern-.125emX}}

\newtheorem{definition}{Definition}
\newtheorem{lemma}{Lemma}
\newtheorem{theorem}{Theorem}

\newtheorem{corollary}{Corollary}
\newtheorem{assumption}{Assumption}
\begin{document}

\title{A $\mu$-Analysis and Synthesis Framework for Partial Integral Equations using IQCs}


\author{Thijs Lenssen, Aleksandr Talitckii, Matthew Peet and Amritam Das
\thanks{T. Lenssen and A. Das are with the Control Systems Group, Eindhoven University of Technology, The Netherlands. A. Talitckii and M. Peet are with School for the Engineering of Matter, Transport and Energy, Arizona State University, the USA. Email addresses: \small{\tt\small{t.m.lenssen@student.tue.nl,atalitck@asu.edu, mpeet@asu.edu,am.das@tue.nl}}}

\thanks{Corresponding author: am.das@tue.nl}}%

\maketitle

\begin{abstract}
We develop a $\mu$-analysis and synthesis framework for infinite-dimensional systems that leverages the Integral Quadratic Constraints (IQCs) to compute the structured singular value's upper bound. The methodology formulates robust stability and performance conditions jointly as Linear Partial Integral Inequalities within the Partial Integral Equation framework, establishing connections between IQC multipliers and $\mu$-theory. Computational implementation via PIETOOLS enables computational tools that practically applicable to spatially distributed infinite dimensional systems. Illustrations with the help of Partial and Delay Differential Equations validate the effectiveness of the framework, showing a significant reduction in conservatism compared to unstructured methods and providing systematic tools for stability-performance trade-off analysis.
\end{abstract}


\section{Introduction}
The integral quadratic constraints (IQC) method, pioneered by Megretski and Rantzer~\cite{megretski1997system}, has its roots in several foundational theories from the 1960s, including absolute stability~\cite{desoer1975feedback}, input-output stability, and dissipative systems theory~\cite{willems1972dissipative1,willems1972dissipative2}.
The absolute stability problem examines feedback systems combining a linear time-invariant component with static nonlinearities characterized by sector bounds. Popov, Yakubovich and Kalman's work establishes crucial connections between stability conditions and linear matrix inequalities (LMIs), which proves valuable for analysis and synthesis. The input-output stability framework, on the other hand, provides powerful tools like the small-gain and passivity theorems. A significant innovation is the introduction of multipliers~\cite{zames1966input}---artificial transfer functions satisfying specific conditions---to reduce conservatism in stability and performance analysis. 

During the 1980's robust control era, attention shifted toward analyzing multi-input multi-output linear time-invariant systems facing structured dynamic and parametric uncertainties ~\cite{fan1991robustness}. The robustness notion refers to maintaining stability and performance across all possible realizations of uncertainty. Early connections between robust stability analysis and multiplier theory, particularly the circle criterion~\cite{safonov1981multiloop} remains the cornerstone of modern control theory. Additionally, thanks to $\mu$-theory~\cite{packard1993complex}, the robust stability problem amounts to computing the structured singular value which became the bedrock of the field of robust performance synthesis. The $\mu$-theory provides complete characterization of the uncertain plant in both time and frequency domain while being tractable by evaluating upper bounds using so called $(D, DG)$ scaling matrices that ultimately serves as multipliers. During the mid-1990s, major advances were observed in optimization theory that enabled efficient computational tools for control system analysis and synthesis based on Linear Matrix Inequalities. Since IQC multipliers are efficiently computed using LMI optimization, the problem of verifying robust stability for diverse uncertainty classes and computing performance bounds using measures derived from dissipative systems theory~\cite{willems1972dissipative1,willems1972dissipative2} has become computationally efficient.

While the framework in~\cite{veenman2016robust} provides comprehensive IQC tools for ODE systems, recent work has extended the IQC methodology to infinite-dimensional systems described by Partial Integral Equations (PIEs) (see \cite{9303892}, \cite{10156335}). The main advantages of PIEs are that a) they are equivalent representations of a wider class of coupled Ordinary Differential Equation (ODE) and Partial Differential Equation (PDE) models and b) they provide a systematic LMI based computational approach for analysis and synthesis \cite{shivakumar2024GDPE}.

Taking inspiration from~\cite{veenman2016robust}, this paper provides practitioners with a systematic tool for analyzing stability and performance of  dynamical systems in the presence of uncertainties where one striking difference is the fact that the dynamics as well as the uncertainties can be equipped with infinite dimensional signals. This paper particularly extends \cite{9303892} and \cite{10156335} in terms of the following contributions:
\begin{enumerate}
\item \textbf{IQC Based Infinite Dimensional Analysis of Robust Stability and Performance:} We derive a single condition that jointly characterizes the robust stability and performance of infinite dimensional operators subject to infinite dimensional uncertainties. We show that, when the operators are chosen to be represented by PIEs, the resulting stability and performance test yields a Linear Partial Integral Inequality (LPI) which can be solved using LMIs. 

\item \textbf{$\mathbf{\mu}$-Analysis:} For the first time, we establish a $\mu$-analysis theory for infinite dimensional systems that establishes properties analogous to those in the finite-dimensional case. We are able to compute upper bounds on the structured singular value for uncertain PIEs with norm-bounded uncertainties.
\item \textbf{Robust Synthesis Applications:} We demonstrate analysis and synthesis capabilities of the proposed robust framework by showcasing stability-performance trade-offs in performance analysis and robust synthesis of Luenberger observer gains.
\end{enumerate}

The remainder of this paper is organized as follows. After some preliminaries in Section II, Section III develops the IQC-based framework for infinite dimensional systems and presents the main results while connecting this framework to $\mu$-theory and structured uncertainty analysis. Section IV details the application of the framework for PIE systems. Section V presents numerical examples validating the approach for both PDE and Delay systems. Section VI concludes with discussion of results and future directions.

\section{Preliminaries}
\subsection{Notation}
$L_2^n[a,b]$ denotes the set of $\R^n$-valued, Lebesgue square-integrable equivalence classes of functions on the spatial domain $[a,b]\subset\R$. The space $\R L_2^{\mathbf{n}}[a,b]$ denotes the Cartesian product $\R^{n_1} \times L_2^{n_2}[a,b]$, where $\mathbf{n} = \bmat{n_1 \\ n_2}$, with inner product
\begin{equation*}
    \ip{\bmat{x_1 \\ \mathbf{x}_2}}{\bmat{y_1 \\ \mathbf{y}_2}}_{\R L_2}
    := x_1^\top y_1 + \ip{\mathbf{x}_2}{\mathbf{y}_2}_{L_2}.
\end{equation*} When clear from context, we occasionally omit the subscript and simply write $\ip{\cdot}{\cdot} := \ip{\cdot}{\cdot}_{\R L_2}$. Since the only spatial domain considered here is $[a,b]$, we typically omit the domain and write $L_2^n$ or $\R L_2^{\mathbf{n}}$, further omitting dimensions when clear from context. For a normed space $X$ and Banach space $Y$, $\mcl L(X,Y)$ denotes the Banach space of bounded linear operators from $X$ to $Y$ with induced norm
\[
\|\mcl P\|_{\mcl L(X,Y)} = \sup_{\|x\| = 1} \|\mcl P \mathbf{x}\|_Y,
\] and $\mcl L(X) := \mcl L(X,X)$. Our notational convention is to write functions in \textbf{bold} (e.g., $\mathbf{x}$, indicating $\mathbf{x} \in \RL$) and operators in calligraphic font (e.g., $\mcl P \in \mcl L(\RL)$).
For a Hilbert space $X$ and operator $\mcl A \in \mcl L(X)$, the adjoint $\mcl A^*$ is defined by
\[
\ip{\mathbf{x}}{\mcl A \mathbf{y}}_X = \ip{\mcl A^* \mathbf{x}}{\mathbf{y}}_X,
\quad \forall\, \mathbf{x},\mathbf{y} \in X.
\] The space $L_2^n[0,\infty)$ denotes $\R^n$-valued square-integrable signals defined on the temporal domain $[0,\infty)$, so that $x(t) \in \R^n$. Similarly, $\RL^n[0,\infty)$ denotes $\RL$-valued square-integrable temporal signals. For suitably differentiable $\mathbf{x} \in \RL^n[0,\infty)$, we write $\mathbf{\dot{x}} := \frac{\partial \mathbf{x}}{\partial t}$. For a Hilbert space $\mathbf{H}$, we denote by $\mathbf{L}_{\mathbf{H}}$ the space of square-integrable functions $u : [0,\infty) \to \mathbf{H}$ with inner product
\[
\ip{\mathbf{u}}{\mathbf{v}}_{\mathbf{L}_{\mathbf{H}}}
= \int_0^\infty \ip{\mathbf{u}(t)}{\mathbf{v}(t)}_{\mathbf{H}}\, dt.
\] We define the extended space $\mathbf{L}_{e,\mathbf{H}}$ to consist of functions $u : [0,\infty) \to \mathbf{H}$ such that
\[
\int_0^T \|\mathbf{u}(t)\|_{\mathbf{H}}^2\, dt < \infty
\quad \forall\, T \ge 0.
\] We use $\mathbf{L}_{e,[a,b]}$ and $\mathbf{L}_{[a,b]}$ to denote $\mathbf{L}_{e,\mathbf{H}}$ and $\mathbf{L}_{\mathbf{H}}$ with $\mathbf{H} = \R L_2[a,b]$. For any Hilbert space $\mathbf{H}$, the truncation operator $p_\tau : \mathbf{L}_{e,\mathbf{H}} \to \mathbf{L}_{e,\mathbf{H}}$ is defined for $\mathbf{y} \in \mathbf{L}_{e,\mathbf{H}}$ by
\begin{equation*}
(p_\tau \mathbf{y})(t) =
\begin{cases}
\mathbf{y}(t), & 0 \le t \le \tau, \\
0, & \text{otherwise}.
\end{cases}
\end{equation*}
\begin{definition}
Let $\mathbf{H}$ and $\mathbf{G}$ be Hilbert spaces. For all $\tau\geq0$, an operator $G : \mathbf{L}_{\mathbf{H}} \to \mathbf{L}_{\mathbf{G}}$ is:
\begin{itemize}
    \item \textbf{Causal} if $p_\tau G = p_\tau G p_\tau$.
    \item \textbf{Bounded on $\mathbf{L}_{\mathbf{H}}$} if $\|G \mathbf{v}\|_{\mathbf{L}_{\mathbf{G}}} \le C \|\mathbf{v}\|_{\mathbf{L}_{\mathbf{H}}}$ for some $C \ge 0$ and all $\mathbf{v} \in \mathbf{L}_{\mathbf{H}}$.
    \item \textbf{Bounded on $\mathbf{L}_{e,\mathbf{H}}$} if $\|p_\tau G \mathbf{v}\|_{\mathbf{L}_{\mathbf{G}}} \le C \|p_\tau\mathbf{v}\|_{\mathbf{L}_{\mathbf{H}}}$ for some $C \ge 0$ and all $\mathbf{v} \in \mathbf{L}_{e,\mathbf{H}}$.
\end{itemize}
\end{definition} 
Finally, for a given $\mcl V \in \mcl L(\RL^\mbf{n}, \RL^\mbf{m})$, we define the operator $\mcl M_{\mcl V} : \mathbf{L}_{e,[a,b]}^\mbf{n} \to \mathbf{L}_{e,[a,b]}^\mbf{m}$ to compactly represent a multiplication operator. For $\mathbf{u} = \bmat{u_1 & \cdots & u_N}^\top$ with $\mathbf{u} \in \RL$, its action at time $t$ is given by the block structure of $\mcl V_{ij}$:
\begin{equation}
\label{eq:block_operator_def}
(\mcl M_{\mcl V} \mathbf{u})(t)
:= \bmat{
\mcl V_{11} & \cdots & \mcl V_{1N} \\
\vdots      & \ddots & \vdots      \\
\mcl V_{N1} & \cdots & \mcl V_{NN}
}
\bmat{u_1 \\ \vdots \\ u_N}(t).
\end{equation}

\subsection{PI operators}
Partial Integral operators, also known as PI operators, are a class of bounded linear integral operators that are defined jointly on a vector space and Hilbert space.
\begin{definition}
    We say $\mcl P \in \mbf{\Pi}_4 \subset \mcl L(\RL^\mbf{n}, \RL^\mbf{m})$, denoted by $\prod \begin{bmatrix} 
    P & Q_1 \\
    Q_2 & \{R\} \\
\end{bmatrix}$, if there exist a matrix $P$ and matrix polynomials $Q_1, Q_2, R_0, R_1$, and $R_2$ (of compatible dimension) such that,
    \begin{gather*}
\Bigg[\Bigg(\prod \begin{bmatrix} 
    P & Q_1 \\
    Q_2 & \{R\} \\
\end{bmatrix}\Bigg)\! \! \bmat{x \\ \mbf{x}}\Bigg](s)\! :=\! \bmat{Px\! +\!\int\limits_a^b\!Q_1(s)\mbf x(s)ds\\Q_2(s)x\!+\!\mcl R \mbf x(s)},\\ 
(\mcl R \mbf x)(s)\!\! =\!\! R_0(s)\mbf x(s)\! +\! \int\limits_a^s\!\!\!R_1(s,\theta)\mbf x(\theta)d\theta\! +\!\int\limits_s^b\!\!\!R_2(s,\theta)\mbf x(\theta) d\theta.
\end{gather*}

\end{definition}
The vector space of PI operators, given compatible dimensions, are closed under composition, addition, adjoint and concatenation -- implying that $\mbf{\Pi}_4$ is a composition algebra \cite{shivakumar2024GDPE}.

\subsection{Linear PI Inequalities (LPIs)}
\begin{definition}(LPIs)
	\label{def_LPI}
For given PI operators $\{\mcl E_{i,j}, \mcl{F}_{i,j}, \mcl{G}_i\}$ and a convex linear functional $\mcl L(\cdot)$, a linear PI inequality is a convex optimization of the following form
\begin{align} 
&\min\limits_{P_i, Q_{1i}, Q_{2i}, R_{0i}, R_{1i}, R_{2i}}  \mcl L(\{P_i, Q_{1i}, Q_{2i}, R_{0i}, R_{1i}, R_{2i}\})\notag\\
&\mathrm{s. t. }\ \sum\limits_{j = 1}^K \mcl E_{ij}^*\prod \begin{bmatrix} 
    P_i & Q_{1i} \\
    Q_{2i} & \{R_i\} \\
\end{bmatrix}\mcl F_{ij} +\mcl G_i \succeq 0
\end{align}
\end{definition}
Only LMIs are required to solve an optimization problem involving LPIs. 
\subsection{PIETOOLS}
 PIETOOLS is a MATLAB toolbox used for declaration, manipulation and solving LPIs \cite{9147712}. For learning how to use PIETOOLS and its functionalities visit \href{https://control.asu.edu/pietools/pietools}{PIETOOLS homepage}.

\section{IQC-Based framework for robust stabilityand performance analysis}
This section develops a comprehensive framework for analyzing stability and performance of interconnected systems in a Hilbert space, utilizing IQCs as the primary analytical tool. We begin by establishing the mathematical foundation for system interconnections and formalizing the key concepts of stability and performance. The IQC methodology is then systematically developed to characterize system uncertainties and derive rigorous conditions that guarantee robust stability and performance. A central contribution of this work lies in bridging IQC theory with $\mu$-analysis for infinite dimensional systems, creating a framework for quantifying their robust stability and performance when those systems are subjected to structured uncertainties. 
\subsection{Description of uncertain systems in a Hilbert-space}
We start by denoting the uncertainty operators as $\Delta$ that belongs to a structured set $\mathbf{\Delta}$ which satisfies the following fundamental property:
\begin{assumption}
    $\Delta\in\mbf{\Delta}$ implies that $\tau\Delta\in\mbf{\Delta}$ for all $\tau \in [0,1]$.
\end{assumption}
This scaling property is essential for developing the IQC-based stability conditions. We consider operators that map external inputs to external outputs through a feedback interconnection between a linear (infinite dimensional) operator $G$ and $\Delta$. This interconnection structure is formally defined as follows
\begin{definition}\label{def1}
    Given operators $G_\Delta : \mbf L_{e,[a,b]}^{\mbf{n}_{w_\Delta}} \to \mbf L_{e,[a,b]}^{\mbf{n}_{z_\Delta}}$, $G_{\Delta w} : \mbf L_{e,[a,b]}^{\mbf{n}_{w_\Delta}} \to L_2^{n_w}[0,\infty)$, $G_{z\Delta } : L_2^{n_z}[0,\infty) \to \mbf L_{e,[a,b]}^{\mbf{n}_{z_\Delta}}$, $G_{zw} : L_2^{n_z}[0,\infty) \to L_2^{n_w}[0,\infty)$, and $\Delta : \mbf L_{e,[a,b]}^{\mbf{n}_{z_\Delta}} \to \mbf L_{e,[a,b]}^{\mbf{n}_{w_\Delta}}$, we define an \textbf{interconnected system} $z = [G,\Delta] (w)$, for inputs $w \in L_2^{n_w}[0,\infty)$ and outputs $z\in L_{2,e}^{n_z}[0,\infty)$, such that the signals $z_\Delta \in \mbf L_{e,[a,b]}^{\mbf{n}_{z_\Delta}}$, $w_\Delta \in \mbf L_{e,[a,b]}^{\mbf{n}_{w_\Delta}}$ satisfy the following relations
    \begin{align}\label{eq:mappingOp}
        z_\Delta &= G_\Delta w_\Delta + G_{\Delta w} w,\\ \notag
        z & = G_{z\Delta} w_\Delta + G_{zw} w, \\
        w_\Delta &= \Delta z_\Delta. \notag
    \end{align}
\end{definition}
The interconnection structure described above captures the feedback interaction between the nominal system $G$ and the uncertainty $\Delta$, where $w$ and $z$ represent external inputs and performance outputs, respectively, while $w_\Delta$ and $z_\Delta$ form the internal feedback loop through the uncertainty. Before analyzing stability and performance, we must first ensure the well-posedness of the interconnection, such that solutions exist, are unique, and depend causally on inputs.
\begin{definition}
  We say that the uncertain interconnection $[G,\Delta]$ is \textbf{well--posed} if:
    \begin{itemize}
        \item \textbf{Existence and Uniqueness}: For any $w\in L_2[0,\infty)$, there exist unique $z\in L_{2,e}[0,\infty)$, $z_\Delta \in \mbf L_{e, [a, b]}$ and $w_\Delta \in  \mbf L_{e, [a, b]}$ such that, $z$, $z_\Delta, w_\Delta$ satisfy the interconnection.
        \item \textbf{Causality}: If $z_\Delta, w_\Delta$ and $\hat{z}_\Delta, \hat{w}_\Delta$ satisfy the interconnection, then $p_\tau(z_\Delta - \hat{z}_\Delta) = 0$ and $p_\tau(w_\Delta - \hat{w}_\Delta) = 0$, for all $\tau \geq 0$.
    \end{itemize}
\end{definition} 
Having established the class of uncertain systems and their well-posedness, we now develop a framework to guarantee robust stability and performance for the class of systems defined in Definition \ref{def1}.
\subsection{Robust stability and performance}
Traditional control system design often treats stability and performance shaping as separate objectives with distinct constraints. However, Integral Quadratic Constraints (IQCs) provide a unified framework where these trade-offs can be captured within a single constraint \cite{10156335}. We now define what it means for the system to remain stable despite uncertainties. Robust stability requires that the system is well-posed for all admissible uncertainties and that the input-output map remains bounded. This involves removal of the performance channels in \eqref{eq:mappingOp}, and introducing auxiliary signals $z_0$ and $w_0$.
\begin{definition}
    The interconnection
    \begin{align*}
        z_\Delta = G_\Delta w_\Delta + z_0, \quad
        w_\Delta = \Delta z_\Delta + w_0, 
    \end{align*}
    with $z_0, w_0 \in \mbf{L}_{e,[a,b]}$, is said to be \textbf{robustly stable} if:
\begin{enumerate}
    \item The uncertain system is well-posed for all $\Delta\in\mbf{\Delta}$.
    \item The map from $\bmat{z_0\\w_0}$ to $\bmat{z_\Delta\\w_\Delta}$ is bounded on $\mbf L_{[a,b]}$ for all $\Delta\in\mbf{\Delta}$.
\end{enumerate}
\end{definition} 
Within the IQC framework, the stability of an interconnection $[G,\Delta]$ is certified by verifying that the uncertainty operator $\Delta$ satisfies a quadratic constraint defined by an appropriate multiplier. This leads to the following formal definition
\begin{definition}\label{def:multipliers}
An uncertainty operator $\Delta: \mbf L_{e,[a,b]}^{\mbf{n}_{z_\Delta}} \to \mbf L_{e,[a,b]}^{\mbf{n}_{w_\Delta}}$ is said to \textbf{satisfy the IQC defined by multiplier $\mcl M_{\mcl V}\in \mcl L(\mbf L_{[a, b]})$}, parameterized by $\mcl{V}\in \mcl L(\RL, \RL)$ if the inequality
\begin{gather*}
    \int_0^T \ip{\bmat{z_\Delta \\ \Delta z_\Delta}(t)}{\bmat{\mcl{V}_{11}&\mcl{V}_{12}\\ \mcl{V}_{21} & \mcl{V}_{22}}\bmat{z_\Delta \\ \Delta z_\Delta}(t)}_{\R L_2} dt \geq 0
\end{gather*}
holds for all $z_\Delta \in \mbf L_{e, [a, b]}$ and all $T \geq 0$.
\end{definition}
Beyond stability, we are interested in ensuring that the system meets certain performance specifications in the presence of uncertainty. Robust performance extends the notion of robust stability by additionally requiring that a performance metric is satisfied.
\begin{definition}\label{def:robperf}
    \textbf{Robust performance} is achieved if the interconnection $[G,\Delta]$ is \textbf{robustly stable} for some performance multiplier $\mcl M_{\hat{\mcl V}}\in \mcl L(L_2)$, parameterized by $\hat{\mcl{V}}\in \mcl L(\R, \R)$ with $\hat{\mcl V}_{11} \succeq 0$ such that
    
    \begin{gather*}\int_0^T\ip{\bmat{([G,\Delta]w) \\ w}(t)}{\bmat{\hat{\mcl{V}}_{11}&\hat{\mcl{V}}_{12}\\\hat{\mcl{V}}_{21}&\hat{\mcl{V}}_{22}}\bmat{([G,\Delta]w) \\ w}(t)}_{\R} dt \\\leq -\epsilon \|p_Tw\|_{L_2}^2
    \end{gather*}
    for all $w \in L_2([0, \infty))$,  all $T \geq 0$, and some $\epsilon >0$.
\end{definition}
In this framework, the multipliers $\mcl M_{\mcl V}$ and $\mcl M_{\hat{\mcl{V}}}$ serve distinct but complementary roles. The multiplier $\mcl M_{\mcl V}$ characterizes the behavior of uncertainties, nonlinearities, and other perturbations captured by the operator $\Delta$, ensuring that the IQC is satisfied for the class of admissible disturbances. In contrast, the multiplier $\mcl M_{\hat{\mcl{V}}}$ encodes performance requirements for the nominal system, such as $L_2$-gain bounds, passivity conditions, or other desired input-output properties. Together, these multipliers enable simultaneous verification of robust stability against perturbations and achievement of specified performance objectives. 

\subsection{Theorem for joint robust stability and performance}

We now present the main theoretical result that provides conditions for simultaneous robust stability and performance verification.

\begin{theorem}[Robust Stability and Performance]\label{thm:metrics}
Let $G$, $\Delta\in\mbf{\Delta}$ be given, and there exists $\mcl{M}_\mcl{V}$, such that,
\begin{itemize}
    \item The interconnection $[G,\Delta]$ is well--posed for all $\Delta\in\mbf{\Delta}$
    \item The IQC in Definition \ref{def:multipliers} is satisfied for all $\Delta\in\mbf{\Delta}$.
\end{itemize}
Then the interconnection $[G,\Delta]$ is \textbf{robustly stable} and \textbf{robust performance} is achieved with respect to some $\mcl{M}_{\hat{\mcl{V}}}$, with $\hat{\mcl V}_{11} \succeq 0$, if,
\begin{equation}\label{eq:rprs}
\begin{gathered}
    \ip{p_T\hat{G}\bmat{w_\Delta\\w}}{p_T\bmat{\mcl{M}_\mcl{V} & 0 \\ 0 & \mcl{M}_{\hat{\mcl V}}}\hat{G}\bmat{w_\Delta\\w}}_{\mbf L_{[a, b]}}\\ \leq -\epsilon (\|p_T{w_\Delta}\|^2_{\mbf L_{[a, b]}}+\|p_T w\|^2_{L_2})
\end{gathered}
\end{equation}
for all $w_\Delta \in \mbf L_{e,[a,b]}^{\mbf{n}_{w_\Delta}}$, $w\in L_2^{n_w}([0, \infty))$ and $T \geq 0$, where $\hat{G} := \left[\begin{array}{cc}
 G_\Delta & G_{\Delta w}\\ I & 0 \\ \hdashline G_{z\Delta} & G_{zw} \\ 0 & I
\end{array}\right]$ according to \eqref{eq:mappingOp}.
\end{theorem}
\begin{proof}
    Suppose that \eqref{eq:rprs} holds. This implies,
        \begin{align*}
        &\ip{p_T \bmat{G_\Delta \\ I}w_\Delta}{p_T\mcl{M}_\mcl{V}\bmat{G_\Delta \\ I}w_\Delta}_{\mbf L_{[a, b]}}\\ &+ \ip{p_T \bmat{G_{z\Delta} \\ 0}w_\Delta}{p_T\mcl{M}_\mcl{\hat{V}}\bmat{G_{z\Delta} \\ 0}w_\Delta}_{\mbf L_{[a, b]}}\\ &\leq -\epsilon \|p_Tw_\Delta\|^2_{\mbf L_{[a, b]}}
        \end{align*}
    Since $\mcl{\hat{V}}_{11}\succeq0$, we can then infer,
    \begin{align*}
        &\ip{p_T \bmat{G_\Delta  w_\Delta\\ w_\Delta}}{p_T\mcl{M}_\mcl{V}\bmat{G_\Delta w_\Delta \\ w_\Delta}}_{\mbf L_{[a, b]}}\leq -\epsilon \|p_Tw_\Delta\|^2_{\mbf L_{[a, b]}}
    \end{align*}
    Then, using Theorem 6 in \cite{10156335}, we find that \textbf{robust stability} is guaranteed with respect to performance multiplier $\mcl{M}_{\hat{V}}$. Moreover, rewriting \eqref{eq:rprs} as,
    \begin{align*}
        &\ip{p_T\bmat{z_\Delta\\w_\Delta}}{p_T\mcl{M}_{{\mcl{V}}}\bmat{z_\Delta \\ w_\Delta}}_{\mbf L_{[a, b]}} + \\&\ip{p_T\bmat{z\\w}}{p_T\mcl{M}_{\hat{\mcl{V}}}\bmat{z \\ w}}_{\mbf L_{[a, b]}}
        \!\!\!\!\!\!\!\!\!\leq  -\epsilon (\|p_T{w_\Delta}\|^2_{\mbf L_{[a, b]}}+\|p_T w\|^2_{L_2}),
    \end{align*}
    and given that $\mcl{M}_\mcl{V}$ satisfies the IQC in Definition \ref{def:multipliers} leads to, 
    \begin{align*}
        \ip{p_T\bmat{z\\w}}{p_T\mcl{M}_{\hat{\mcl{V}}}\bmat{z \\ w}}_{\mbf L_{[a, b]}}\!\!\!\!\!\!\!\!\!\leq  -\epsilon \|p_T w\|^2_{L_2}.
    \end{align*}
    Thus by Definition \ref{def:robperf} \textbf{robust performance} is achieved.
\end{proof}

Theorem \ref{thm:metrics} establishes a framework for deriving Linear PI Inequalities (LPIs), which are used in assessing the robust stability of a system. This result can be used to establish $\mu$-analysis tools for infinite dimensional systems.

\subsection{Connection to $\mu$-theory}\label{sec:mu}

Practical control systems contain uncertainties whose structure, when exploited, enables less conservative robustness analysis than unstructured approaches. Real-world uncertainties often exhibit patterns or constraints that structured uncertainty sets can capture. The $\mu$-theory \cite{packard1993complex} provides a foundational framework for norm-bounded structured uncertainties in finite dimensions. To extend similar concepts to infinite dimension, we consider Partial Integral operators that are augmented in a block-diagonal fashion yielding the following uncertainty operator:
\begin{align}
    \Delta_\mu = \text{diag}(\delta_1 I, \cdots, &\delta_p I, \delta_{p+1}(t)I, \cdots, \delta_{p+q}(t)I,\notag\\& \Delta_{p+q+1}, \cdots, \Delta_{p+q+r}) \in \mbf{\Pi}_4 \label{eq1}
\end{align}
where
\begin{enumerate}
    \item $\delta_i\in\R$, $i\in[1,p]$ are static real parameters with $|\delta_i|<1$.
    \item $\delta_j(t)\in\R$, $j\in[p+1,p+q]$ are real, time-varying parameters with $\sup_{t>0}|\delta_j(t)|<1$.
    \item $\Delta_k \in \mbf{\Pi}_4$, $k\in[p+q+1,p+q+r]$ are PI operators with $\|\Delta_k\|_{\RL}<1$.
\end{enumerate}
Let the set of all such $\Delta_\mu$ be denoted by $\mbf{\Delta}_\mu$. Since, in the case of finite dimensional systems, determining the robust stability margin requires verifying the existence and stability of $(I - G_{\Delta} \Delta_{\mu})^{-1}$ which can be reduced to computing the upper-bound of the structured singular value. However, in case of infinite dimensional systems, such computations are not always possible. Here, we adopt a definition of structured singular values based on robust stability.

\begin{definition}[Structured singular value]
Given a linear system \( H \) and a norm-bounded structured uncertainty set \( \mbf{\Delta} \), the \emph{structured singular value} \( \mu_{\Delta}(H) \) is the infimum of all \( \gamma \) for which some interconnection of the form,
\begin{align*}
    q = Hp, \quad p=\Delta q,
\end{align*}
remains robustly stable for all scaled uncertainties in the set $\mbf{\Delta}$, i.e.,
\begin{equation*}
\mu_{\Delta}(H) = \inf \left\{ \gamma \geq 0 \;\middle|\; \begin{aligned}&\text{The interconnection} \\&\text{ is stable } \forall \Delta \in \tfrac{1}{\gamma}\mbf{\Delta}\end{aligned} \right\}.
\end{equation*}
\end{definition}
This notion of structured singular value provides a direct criterion for robust stability: the interconnection is robustly stable for all uncertainties $\Delta \in \mathbf{\Delta}$ if
\begin{equation*}
    \mu_{\Delta}(H) \leq 1.
\end{equation*}
To incorporate performance objectives, we now introduce an auxiliary uncertainty operator $\hat{\Delta}$ that encodes the performance specifications on the $z$-$w$ channel. This allows us to formulate an extended uncertainty structure
\begin{equation*}
    \Delta_e = \begin{bmatrix}\Delta_\mu & 0 \\ 0 & \hat{\Delta}\end{bmatrix},
\end{equation*}
which simultaneously captures both the plant uncertainties and performance requirements. Within this framework, we can draw the following conclusions:
\begin{itemize}
    \item \textbf{Robust Stability}: If $\mu_{\Delta_\mu}(G) \leq 1$, then the interconnection $[G,\Delta_\mu]$ is robustly stable, ensuring well-posedness and boundedness of the map from $w_\Delta$ to $z_\Delta$ for all $\Delta_\mu \in \mathbf{\Delta}_\mu$. \textbf{Note:} For stability analysis set $w=z=0$ in \eqref{eq:mappingOp}.
    \item \textbf{Robust Performance}: If $\mu_{\Delta_e}(G) \leq 1$, then the system achieves robust performance. This incorporates the performance constraint $w = \hat{\Delta}z$ into the interconnection relations defined in \eqref{eq:mappingOp}, guaranteeing that both stability and performance specifications are met simultaneously.
\end{itemize}
\begin{figure}
    \centering
    \includegraphics[width=0.65\linewidth]{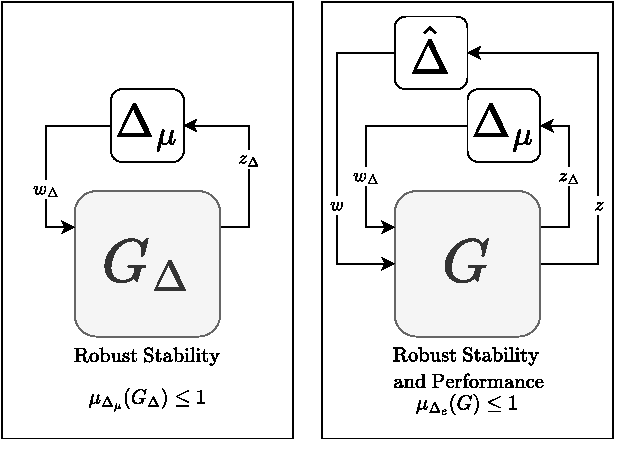}
    \caption{Block diagram interpretations of robust stability (left) and robust performance (right) in $\mu$-theory. The robust performance case includes the performance operator $\hat{\Delta}$ that enforces specifications on the $z$-$w$ channel.}
    \label{fig:RSandRP}
\end{figure}
Figure~\ref{fig:RSandRP} illustrates the system interconnections and summarizes the conditions for robust stability and performance within the $\mu$-framework. Although the structured singular value provides a rigorous theoretical foundation, computing $\mu_\Delta(G)$ exactly remains challenging, even for finite-dimensional systems where bounds are typically employed. For infinite-dimensional systems, the IQC framework offers a computationally viable alternative by enabling the computation of guaranteed upper bounds on $\mu_\Delta(G)$. 

\subsection{Usage of multipliers}
To bridge $\mu$-theory with the IQC framework, we construct multipliers that characterize the scaled uncertainty set $\tfrac{1}{\gamma_V}\mathbf{\Delta}_\mu$, for a $\gamma_V$, ensuring that every $\Delta_\mu \in \tfrac{1}{\gamma_V}\mathbf{\Delta}_\mu$ satisfies the corresponding IQCs. Exploiting the fact that the choice of uncertainties cover PI operators according to \eqref{eq1}, the multiplier construction proceeds as follows
\begin{itemize}
    \item \textbf{Static real uncertainties} $\delta_i I$ for $|\delta_i| \leq \tfrac{1}{\gamma_V}$, $i \in [1,p]$:
    \begin{equation}
    \label{eq2}
        \mcl{V}_i = \begin{bmatrix}
            \mcl{P}_i & \gamma_V\mcl{R}_i \\
            \gamma_V\mcl{R}_i^* & -\gamma_V^2\mcl{P}_i
        \end{bmatrix}
    \end{equation}
    
    \item \textbf{Time-varying real uncertainties} $\delta_j(t) I$ for $\sup_{t>0}|\delta_j(t)| \leq \tfrac{1}{\gamma_V}$, $j \in [p+1,p+q]$:
    \begin{equation}
    \label{eq3}
        \mcl{V}_j = \begin{bmatrix}
            \mcl{P}_j & \gamma_V\mcl{R}_j \\
            \gamma_V\mcl{R}_j^* & -\gamma_V^2\mcl{P}_j
        \end{bmatrix}
    \end{equation}
    
    \item \textbf{PI operator uncertainties} $\|\Delta_k\|_{\RL} \leq \tfrac{1}{\gamma_V}$ for $k \in [p+q+1,p+q+r]$:
    \begin{equation}
    \label{eq4}
        \mcl{V}_k = \begin{bmatrix}
            I & 0 \\
            0 & -\gamma_V^2I
        \end{bmatrix}
    \end{equation}
\end{itemize}
where $\mathcal{P}_l, \mcl{R}_l \in \mathbf{\Pi}_4$ satisfy $\mcl{R}_l^* = -\mcl{R}_l$ and $\mathcal{P}_l^* = \mathcal{P}_l \succ 0$ for $l\in[1,p+q]$. The validity of these multipliers follows directly from Lemma 10 and Corollary 11 in~\cite{10156335} with $\Psi = I$. 

To address performance objectives, we introduce a complementary performance multiplier. Given the finite-dimensional nature of the performance channels, we restrict our attention to the multiplier classes detailed in~\cite{veenman2016robust}, Section 6. Specifically, we focus on the induced $L_2$-gain multiplier for performance characterization.
\begin{lemma}[Performance Characterization]\label{lem:perf}
    Suppose \textbf{robust performance} is achieved for $[G,\Delta]$ with the performance multiplier $\mcl{M}_{\hat{\mcl{V}}}$ parameterized as
    \begin{equation}\label{eq:perfV}
        \hat{\mcl{V}} = \begin{bmatrix} I & 0 \\ 0 & -\gamma_{\hat{V}}^2I \end{bmatrix},
    \end{equation}
    where $\gamma_{\hat{V}} > 0$. Then the worst-case induced $L_2$-gain satisfies
    \begin{equation*}
        \sup_{\Delta\in\mbf{\Delta}} \|[G,\Delta]\|_\infty \leq \gamma_{\hat{V}}.
    \end{equation*}
\end{lemma}
\begin{proof}
    Suppose robust performance is achieved for the interconnection $[G,\Delta_e]$ using a multiplier $\mcl{M}_{\hat{\mcl{V}}}$ parameterized as above, then 
    \begin{equation*}
        \int_0^T\ip{\bmat{[G,\Delta]w\\w}}{\bmat{I & 0\\ 0 & -\gamma_{\hat{V}}^2I}\bmat{[G,\Delta]w\\w}}dt\leq -\epsilon\|p_Tw\|^2_{L_2},
    \end{equation*}
    gives,
    \begin{equation*}
        \frac{\|p_T [G,\Delta]w\|_{L_2}}{\|p_Tw\|_{L_2}}\leq\gamma_{\hat{V}}.
    \end{equation*}
    Since \textbf{robust performance} implies \textbf{robust stability} we can thus conclude
    $\sup_{\Delta\in\mbf{\Delta}}\|[G,\Delta]\|_\infty\leq\gamma_{\hat{V}}, \text{ as } T\to\infty.$
\end{proof}
The parameters $\gamma_V$ and $\gamma_{\hat{V}}$ enable explicit trade-offs between stability and performance, with the bounds $\mu_{\Delta_\mu}(G) \leq \gamma_V$ and $\mu_{\Delta_e}(G) \leq \max\{\gamma_V, \gamma_{\hat{V}}\}$. This multiplier construction preserves the structured uncertainty interpretation of $\mu$-theory while leveraging the computational advantages of the IQC framework. 
\section{Application to PIE systems}
\begin{figure}[h]
    \centering
    \includegraphics[width=0.75\linewidth]{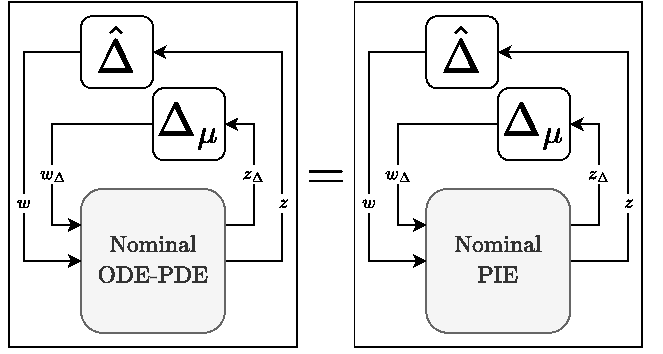}
    \caption{Equivalence between ODE-PDE and PIE Linear Fractional Representations, showing the nominal system interconnected with stability ($\Delta$) and performance ($\hat{\Delta}$) operators.}
    \label{fig:block structure}
\end{figure}

The PIE framework provides systematic methods for converting coupled ODE-PDE systems into equivalent Partial Integral Equation (PIE) representations. 
Therefore, if $G$ in Definition \ref{def1} admits an ODE-PDE based representation, it can be equivalently represented as a PIE. Consequently, as shown in Figure \ref{fig:block structure}, $[G, \Delta]$ can be represented as an interconnection between PIEs and uncertainties. 
\begin{definition}\label{def:uPIE}
Let the PI operators be $\mathcal{T}, \mathcal{A}: \mathcal{L}(\RL^{\mbf{n}_x}), \quad \mathcal{B}_\Delta: \RL^{\mbf{n}_{w_\Delta}} \to \RL^{\mbf{n}_x}, \quad \mathcal{B}_w: \R^{n_w} \to \RL^{\mbf{n}_x}, \quad \mathcal{C}_\Delta: \RL^{\mbf{n}_x} \to \RL^{\mbf{n}_{z_\Delta}}, \quad \mathcal{D}_\Delta: \R^{\mbf{n}_{w_\Delta}} \to \RL^{\mbf{n}_{z_\Delta}}, \quad \mathcal{D}_{\Delta w}: \R^{n_{w}} \to \RL^{\mbf{n}_{z_\Delta}}, \quad , \mathcal{C}_z: \RL^{\mbf{n}_{x}} \to \R^{n_z}, \quad \mathcal{D}_{z\Delta}: \RL^{\mbf{n}_{w_\Delta}} \to \R^{n_{z}}, \quad , \mathcal{D}_{zw} \in \R^{n_z \times n_w}$, with $\mathbf{x}_f \in \RL^{\mbf{n}_x}$, regulated output $z\in L_{2,e}^{n_z}[0,\infty)$, disturbance $w\in L_2^{n_w}[0,\infty)$. For uncertainty $\Delta_\mu \in \mathbf{\Delta}_\mu$, defined according to \eqref{eq1}, construct the following uncertain PIE,
\begin{align}
    \mathcal{T}\dot{\mathbf{x}}_f(t) &= \tilde{\mathcal{A}}(\Delta_{\mu})\mathbf{x}_f(t)+\tilde{\mathcal{B}}_w(\Delta_{\mu})w(t),\notag\\
    z(t) &= \tilde{\mathcal{C}}_z(\Delta_{\mu})\mathbf{x}_f(t)+\tilde{D}_{zw}(\Delta_{\mu})w(t).
    \label{eq:uPIE}
\end{align}
The uncertain PIE \eqref{eq:uPIE} is then defined to admit a $[G, \Delta_{\mu}]$ form according to Definition \ref{def1} if there exist PI operators $\{\mcl T, \mcl A, \mcl B_\Delta, \mcl B_w, \mcl C_\Delta, \mcl D_\Delta, \mcl D_{\Delta w}, \mcl C_z, \mcl D_{z\Delta}, \mcl D_{zw}\}$ that parametrize the following PIEs for all $(\mathbf{x}_f(t), w_\Delta(t), z_\Delta(t), w(t), z(t))$, and for all $\Delta_{\mu}\in\mathbf{\Delta}_{\mu}$,
\begin{equation}
    {G :\left\{\begin{aligned}
        \mathcal{T}\dot{{\mathbf{x}}}_f(t) &= \mathcal{A}{\mathbf{x}}_f(t) + \mathcal{B}_\Delta w_\Delta(t) +\mathcal{B}_w w(t),\notag\\
        z_\Delta(t) &= \mathcal{C}_\Delta\mathbf{x}_f(t) + \mathcal{D}_\Delta w_\Delta(t) + \mathcal{D}_{\Delta w} w(t),\notag\\
        z(t) &= \mathcal{C}_z\mathbf{x}_f(t) + \mathcal{D}_{z\Delta} w_\Delta(t) + \mathcal{D}_{zw} w(t),
    \end{aligned}\right.}
\end{equation}
\begin{center}
   with, $w_\Delta = \Delta_{\mu} z_\Delta.$ 
\end{center}

\end{definition} 

\begin{theorem}\label{thm:robSYN}

Let $G$ be parametrized by PI operators $\{\mcl T, \mcl A, \mcl B_\Delta , \mcl B_w,\mcl C_\Delta, \mcl D_\Delta, \mcl D_{\Delta w}, \mcl C_z, \mcl D_{z\Delta}, \mcl D_{zw}\}$. Let $\Delta_\mu\in\tfrac{1}{\gamma_V}\mbf{\Delta}_\mu$, and there exists a multiplier $\mcl{M}_{\mcl{V}_{\mu}}$ parameterized according to \eqref{eq1}, such that
\begin{itemize}
    \item The interconnection $[G,\Delta_\mu]$ is well--posed for all $\Delta_\mu\in\tfrac{1}{\gamma_V}\mbf{\Delta}_\mu$
    \item The IQC in Definition \ref{def:multipliers} is satisfied for all $\Delta_\mu\in\tfrac{1}{\gamma_V}\mbf{\Delta}_\mu$.
\end{itemize}
Then the interconnection $[G,\Delta_\mu]$ is \textbf{robustly stable}, with $\mathbf{x}_f(0) = 0$ in Definition \ref{def:uPIE},  and \textbf{robust performance} is achieved if there exists some $\mcl{M}_{\hat{\mcl{V}}}$, with $\hat{\mcl V}_{11} \succeq 0$, some $\mathcal{R}$, $\mathcal{Q}\in \Pi_4$ with $\mathcal{R}=\mathcal{R}^*\succeq 0 $, such that,

\begin{itemize}
    \item $\mathcal{T}^* \mathcal{Q} = \mathcal{Q}^* \mathcal{T} = \mathcal{R}$,

    \item and
\vspace{-3ex}
    \begin{align*}
    &\begin{bmatrix}
        \mathcal{A}^* \mathcal{Q} +  \mcl{Q}^*\mcl{A} & \mcl{Q}^*\mcl{B}_\Delta & \mcl{Q}^*\mcl{B}_w\\
        \mathcal{B}_\Delta^* \mathcal{Q} & \epsilon I & 0 \\
        \mathcal{B}_w^* \mathcal{Q}  & 0 & \epsilon I
    \end{bmatrix} \\
    &\quad + \begin{bmatrix}
        \mathcal{C}^*_{\Delta} & 0 \\
        \mathcal{D}^*_{\Delta} & I \\
        \mathcal{D}^*_{\Delta w} & 0
    \end{bmatrix}
        \mathcal{V}_{\mu}
    \begin{bmatrix}
        \mathcal{C}_{\Delta} & \mathcal{D}_{\Delta} & \mathcal{D}_{\Delta w}\\
        0 & I & 0
    \end{bmatrix}\\
    &\quad + \begin{bmatrix}
        \mathcal{C}^*_z & 0 \\
        \mathcal{D}^*_{z\Delta} & 0 \\
        \mathcal{D}^*_{ w} & I
    \end{bmatrix}{\hat{\mcl{V}}}\begin{bmatrix}
        \mathcal{C}_z & \mathcal{D}_{z\Delta} & \mathcal{D}_{z w}\\
        0 & 0 & I
    \end{bmatrix}\preceq 0.
\end{align*}
\end{itemize}
Here, $\mathcal{V}_{\mu}$ is built by diagonally stacking the appropriate multipliers from \eqref{eq2}–\eqref{eq4}, depending on whether $\Delta_{\mu}$ in \eqref{eq1} is static real, time-varying real, or a PI operator.
\end{theorem}
\begin{proof}

Suppose the LPI holds. Define $V(\mbf{x}_f) = \langle\mbf{x}_f, \mcl{R}\mbf{x}_f\rangle \geq 0$. Since $\mcl{Q}^*\mcl{T} = \mcl{T}^*\mcl{Q} = \mcl{R}$, we have $V(\mbf{x}_f) = \langle\mcl{T}\mbf{x}_f, \mcl{Q}\mbf{x}_f\rangle = \langle\mcl{Q}\mbf{x}_f,\mcl{T}\mbf{x}_f\rangle$. Now, for any $z_\Delta \in \RL$, $w_\Delta \in \RL$, $z\in L_2^{n_z}[0,\infty)$, and $w \in L_2^{n_w}[0,\infty)$ satisfying the interconnection $[G,\Delta_{\mu}]$. By the inequality above, we have that,
\begin{align*}
&\dot{V}(\mbf{x}_f(t)) +\epsilon (\|{w_\Delta}(t)\|^2_{\RL}+\| w(t)\|^2_{\R})+
\\&\left\langle
    \begin{bmatrix}
        z_\Delta(t) \\[2pt]
        w_\Delta(t)
    \end{bmatrix},
    \mathcal{V}_{\mu}
    \begin{bmatrix}
        z_\Delta(t) \\[2pt]
        w_\Delta(t)
    \end{bmatrix}
\right\rangle_{\mathbb{R}L_2} + \left\langle
    \begin{bmatrix}
        z(t) \\[2pt]
        w(t)
    \end{bmatrix},
    \hat{\mathcal{V}}
    \begin{bmatrix}
        z(t) \\[2pt]
        w(t)
    \end{bmatrix}
\right\rangle_{\mathbb{R}L_2} \leq 0
\end{align*}
Now, since $V(\mbf{x}_f(0)) = V(0) = 0$ and $V(\mbf{x}_f(T))\geq0$, after integrating in time, we obtain,
\begin{align*}
    &\epsilon (\|{w_\Delta}(t)\|^2_{\RL}+\| w(t)\|^2_{\R})  + \int_0^T \left\langle
    \begin{bmatrix}
        z_\Delta(t) \\[2pt]
        w_\Delta(t)
    \end{bmatrix},
    \mathcal{V}_{\mu}
    \begin{bmatrix}
        z_\Delta(t) \\[2pt]
        w_\Delta(t)
    \end{bmatrix}
\right\rangle_{\mathbb{R}L_2} \\
&+ \left\langle
    \begin{bmatrix}
        z(t) \\[2pt]
        w(t)
    \end{bmatrix},
    \hat{\mathcal{V}}
    \begin{bmatrix}
        z(t) \\[2pt]
        w(t)
    \end{bmatrix}
\right\rangle_{\mathbb{R}L_2} dt\leq -V(\mbf{x}_f(T)) + V(\mbf{x}_f(0))
\end{align*}
We conclude that,
\begin{equation*}
\begin{gathered}
    \ip{p_T\hat{G}\bmat{w_\Delta\\w}}{p_T\bmat{\mcl{M}_{\mcl{V}_{\mu}} & 0 \\ 0 & \mcl{M}_{\hat{\mcl V}}}\hat{G}\bmat{w_\Delta\\w}}_{\mbf L_{[a, b]}}\\ \leq -\epsilon (\|p_T{w_\Delta}\|^2_{\mbf L_{[a, b]}}+\|p_T w\|^2_{L_2}),
\end{gathered}
\end{equation*}
where $\hat{G} := \left[\begin{array}{cc}
 G_\Delta & G_{\Delta w}\\ I & 0 \\ \hdashline G_{z\Delta} & G_{zw} \\ 0 & I
\end{array}\right]$. Then by Theorem 1 we find that robust performance and stability are guaranteed.
\end{proof}
This general LPI framework can be applied to solve robust observer synthesis. Moreover, after establishing a duality result as in \cite{shivaknew} this framework also provides the tools to perform robust control synthesis.
\begin{corollary}[Robust Luenberger Observer Gain Synthesis]\label{cor:robSYN}

Let $G$ be parametrized by PI operators $\{\mcl T, (\mathcal{A}+\mathcal{L}\mathcal{C}_y), (\mathcal{B}_\Delta + \mathcal{L}\mathcal{D}_{y\Delta}), (\mathcal{B}_w+\mathcal{L}\mathcal{D}_{yw}),\mcl C_\Delta, \mcl D_\Delta, \mcl D_{\Delta w}, \mcl C_z, \mcl D_{z\Delta}, \mcl D_{zw}\}$. Let $\Delta_\mu\in\tfrac{1}{\gamma_V}\mbf{\Delta}_\mu$, and there exists a multiplier $\mcl{M}_{\mcl{V}_{\mu}}$ parameterized according to \eqref{eq1}, such that
\begin{itemize}
    \item The interconnection $[G,\Delta_\mu]$ is well--posed for all $\Delta_\mu\in\tfrac{1}{\gamma_V}\mbf{\Delta}_\mu$
    \item The IQC in Definition \ref{def:multipliers} is satisfied for all $\Delta_\mu\in\tfrac{1}{\gamma_V}\mbf{\Delta}_\mu$.
\end{itemize}
Then the interconnection $[G,\Delta_\mu]$ is \textbf{robustly stable}, with $\mathbf{e}_f(0) = 0$,  and \textbf{robust performance} is achieved if there exists some, \(
Z_1 \in \mathbb{R}^{n_x \times n_y}, \quad Z_2 : [a,b] \to \R^{n_p\times n_y},
\) $\mcl{M}_{\hat{\mcl{V}}}$, with $\hat{\mcl V}_{11} \succeq 0$, some $\mathcal{R}$, $\mathcal{Q}\in \Pi_4$ with $\mathcal{R}=\mathcal{R}^*\succeq 0 $, $\mcl{Q}$ is invertible, such that,

\begin{itemize}
\item $\mathcal{Z} := \prod \begin{bmatrix} 
    Z_1 & \emptyset \\
    Z_2 & \{\emptyset\} \\
\end{bmatrix},$
    \item $\mathcal{T}^* \mathcal{Q} = \mathcal{Q}^* \mathcal{T} = \mathcal{R}$,

    \item $\mathcal{L} = \left(\mathcal{Q}^*\right)^{-1}\mathcal{Z}$ ,

    \item and
\vspace{-2ex}
    \begin{align*}
    &\begin{bmatrix}
        (\mathcal{A}^* \mathcal{Q} + \mathcal{C}_y^* \mathcal{Z}^*) + (\cdot)^* & (\cdot)^* & (\cdot)^*\\
        \mathcal{B}_\Delta^* \mathcal{Q} + \mathcal{D}_{y\Delta}^* \mathcal{Z}^* & \epsilon I & 0 \\
        \mathcal{B}_w^* \mathcal{Q} + \mathcal{D}_{yw}^* \mathcal{Z}^* & 0 & \epsilon I
    \end{bmatrix} \\
    &\quad +\begin{bmatrix}
        \mathcal{C}^*_{\Delta} & 0 \\
        \mathcal{D}^*_{\Delta} & I \\
        \mathcal{D}^*_{\Delta w} & 0
    \end{bmatrix}
        \mathcal{V}_{\mu}
    \begin{bmatrix}
        \mathcal{C}_{\Delta} & \mathcal{D}_{\Delta} & \mathcal{D}_{\Delta w}\\
        0 & I & 0
    \end{bmatrix}\\
    &\quad + \begin{bmatrix}
        \mathcal{C}^*_z & 0 \\
        \mathcal{D}^*_{z\Delta} & 0 \\
        \mathcal{D}^*_{ w} & I
    \end{bmatrix}{\hat{\mcl{V}}}\begin{bmatrix}
        \mathcal{C}_z & \mathcal{D}_{z\Delta} & \mathcal{D}_{z w}\\
        0 & 0 & I
    \end{bmatrix}\preceq 0.
\end{align*}
\end{itemize}
Here, $\mathcal{V}_{\mu}$ is built by diagonally stacking the appropriate multipliers from \eqref{eq2}–\eqref{eq4}, depending on whether $\Delta_{\mu}$ in \eqref{eq1} is static real, time-varying real, or a PI operator.
\end{corollary}
\begin{proof}
    After substituting $\mcl{Z} = \mcl{Q}^*\mcl{L}$ in the LPI above, the proof is identical to the one from Theorem~\ref{thm:robSYN}.
\end{proof}

\section{Numerical Examples}
In this section, we present numerical examples that demonstrate the trade-off between robust stability and performance. All examples follow a consistent methodology: we first express the system dynamics in \emph{Linear Fractional Representation (LFR) form} using appropriate perturbation signals, then convert to a PIE representation using  PIETOOLS~\cite{9147712}, and finally analyze the robust stability and performance by solving the LPIs. Throughout our experiments, we employ the 'light' settings of PIETOOLS where the numerical tolerance is chosen as $\epsilon = 10^{-9}$. 
\subsection{Stability of the Diffusion-Reaction Equation}
Consider the heat equation with parametric uncertainty in the diffusion and reaction terms, for \( s \in [0, L] \) and \( t \geq 0 \):
\begin{equation}\label{eq:heat}
\begin{aligned}
    T_t(s, t) &= d(\delta_1) T_{ss}(s, t) + \lambda(\delta_2) T(s, t), \\
    T(0, t) &= 0, \quad T_s(L, t) = 0,
\end{aligned}
\end{equation}
where \( d(\delta_1) = \hat{d} + \tilde{d} \delta_1 \) and \( \lambda(\delta_2) = \hat{\lambda} + \tilde{\lambda} \delta_2 \), with \( \delta_1, \delta_2 \in [-1, 1] \) and \( \tilde{d}, \tilde{\lambda} > 0 \). The uncertainty is thus modeled as \( (\Delta z_\Delta)(t) = \begin{bmatrix} \delta_1 & 0 \\ 0 & \delta_2 \end{bmatrix} \begin{bmatrix} z_{\Delta, 1} \\ z_{\Delta, 2} \end{bmatrix}(t) \).
To express this system in the form of a Linear Fractional Representation (LFR), define the perturbation signals:
\[
w_{\Delta, i}(s, t) = \delta_i z_{\Delta, i}(s, t) \quad \text{for} \quad i = 1, 2.
\]
The system dynamics become:
\begin{align*}
    T_t(s, t) &= \hat{d} T_{ss}(s, t) + \hat{\lambda} T(s, t) + \tilde{d} w_{\Delta, 1}(s, t) + \tilde{\lambda} w_{\Delta, 2}(s, t), \\
    z_{\Delta, 1}(s, t) &= T_{ss}(s, t), \quad z_{\Delta, 2}(s, t) = T(s, t).
\end{align*}
Applying the method of separation of variables to \eqref{eq:heat} yields the stability condition:
\[
\lambda(\delta_2) < d(\delta_1) c, \quad \text{where} \quad c = \left( \frac{\pi}{2L} \right)^2.
\]
The stability condition becomes:
\[
\hat{\lambda} + \frac{\tilde{\lambda}}{\gamma_V} \delta_2 < \left( \hat{d} + \frac{\tilde{d}}{\gamma_V} \delta_1 \right) c.
\]
The worst-case scenario occurs when the uncertainties are maximized on the left-hand side and minimized on the right-hand side, i.e., when \( \delta_1 = -1 \) and \( \delta_2 = 1 \). This leads to the following analytical bound for the maximum allowable uncertainty:
\[
\gamma_V^* = \frac{\tilde{d}c + \tilde{\lambda}}{\hat{d}c - \hat{\lambda}}.
\]

Table~\ref{tab:gamma_comparison} summarizes the stability margins from different methods. The analytical value $\gamma_V^*$ serves as the baseline. The unstructured approach yields the conservative bound $\bar{\gamma}$ by ignoring uncertainty structure, while the bisection method provides the least conservative result $\hat{\gamma}$ by minimizing $\gamma_V$ via LPI-based search.

\begin{table}[htbp]
\centering
\caption{Stability margin comparison for Example~A}
\label{tab:gamma_comparison}
\begin{tabular}{lccc}
\hline
Sample & Analytical & Unstructured   & Bisection \\
       & $\gamma_V^*$ & $\bar{\gamma_V}$ & $\hat{\gamma_V}$ \\\hline
\#1     & $5.2800$   & $5.4229$       & $5.2711$\\\hline
\#2     & $\infty$   & Not feasible   & Not feasible \\ \hline
\#3     & $0.1642$   & $0.1808$       & $0.1636$\\\hline
\#4     & $121.7375$ & $134.3609$     & $121.6659$\\\hline
\hline
\hline
\end{tabular}
\end{table}\vspace{-0.5cm}

\subsection{Delay Differential Equation} 
\begin{figure}[H]
    \centering
    \includegraphics[width=0.7\linewidth]{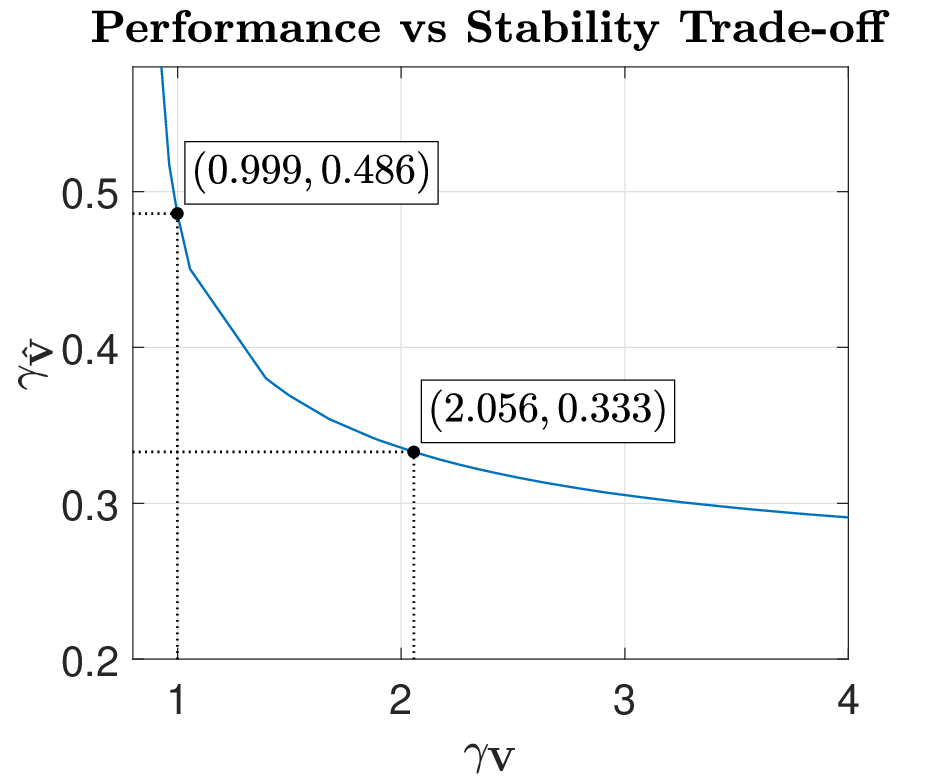}
    \caption{Stability-performance trade-off curve showing the relationship between robustness ($\gamma_V$) and performance ($\gamma_{\hat{V}}$). The curve exhibits asymptotic behavior where performance improvements become minimal despite significant stability compromises.}
    \label{fig:tradeoff}
\end{figure}
In this example, consider the Delay Differential Equation (DDE)~\cite{GuKharitonovChen2003}:
\[
\ddot{y}(t) = 0.1 \dot{y}(t) - 2 y(t) - y(t-\tau) + 0.1w(t), \qquad \tau>0.
\]
Using the delay state \(\phi(t,s) = x(t-\tau s), \, s\in[0,1]\) and ODE state \(x = \begin{bmatrix} x_1 & x_2 \end{bmatrix}^{\top} = \begin{bmatrix} y & \dot y \end{bmatrix}^{\top}\), the DDE dynamics can be converted to ODE--PDE dynamics ~\cite{peet2020representationnetworkssystemsdelay}, then the equations for analysis become:
\begin{figure*}[t]
    \centering
    \includegraphics[width=0.8\linewidth]{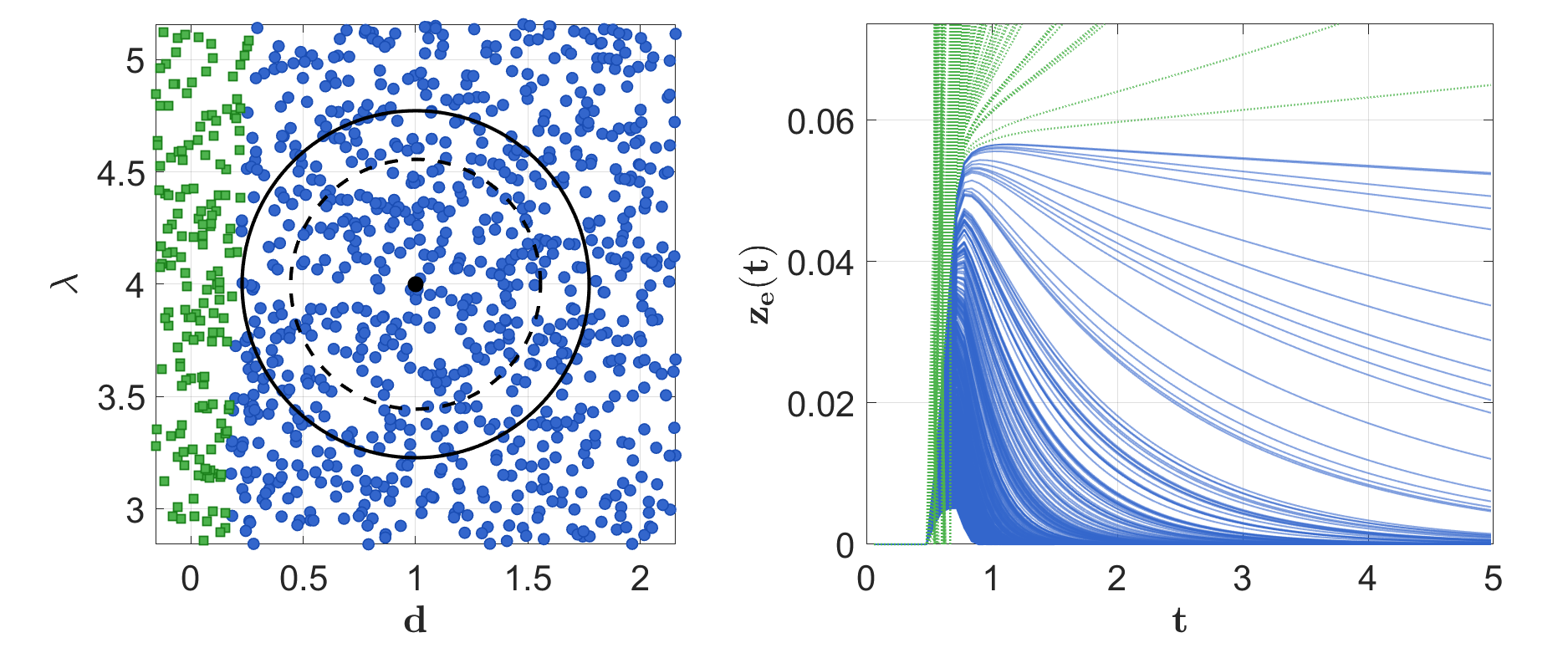}
    \caption{Robust synthesis comparison: (Left) Parameter space with stable (blue) and unstable (green) variables, showing admissible uncertainty sets for structured (solid) and unstructured (dashed) synthesis. (Right) Trajectories simulated using structured-synthesis observer gain, maintaining consistent color coding. At $0.5\leq t \leq 0.75$ a step disturbance was applied to the system.}
    \label{fig:montecarlo}
\end{figure*}

\begin{align*}
\dot{x}(t) &= \begin{bmatrix} x_2(t) \\ 0.1 x_2(t) - 2 x_1(t) - \phi(t,1) + 0.1w(t) \end{bmatrix},\\
\dot{\phi}(t,s) &= -\tfrac{1}{\tau_0} \phi_s(t,s) + w_\Delta(t,s),\\
\phi(t,0) &= x_1(t), \quad z_\Delta(t,s) = \phi_s(t,s), z(t) = \int_0^1 \phi(t,\theta)\,d\theta,
\end{align*}
where $\tau_0 = 0.5$. By defining $w_\Delta(t,s) = \delta z_\Delta(t,s)$ with $\delta\in[-1,1]$. We then create a grid of $\gamma_V$ values and calculate the corresponding $\gamma_{\hat{V}}$ to indicate a trade-off between performance and stability. This curve is shown in Figure~\ref{fig:tradeoff}.

\subsection{Diffusion-Reaction Robust Observer Synthesis}

 Consider the heat equation with parametric uncertainty as in \eqref{eq:heat}:
\[
d(\delta_1) = \hat{d} + \tilde{d}\,\delta_1, \qquad
\lambda(\delta_2) = \hat{\lambda} + \tilde{\lambda}\,\delta_2,
\]
with \(\delta_1, \delta_2 \in [-1,1]\), \(\hat d = \tilde d = 1\), and \(\hat \lambda = 4, \tilde \lambda = 1\). Define the perturbation signals
\[
w_{\Delta,i}(s,t) = \delta_i\, z_{\Delta,i}(s,t), \quad i = 1,2.
\]
The dynamics then become
\begin{align*}
&T_t(s,t) = \hat{d} T_{ss}(s,t) + \hat{\lambda} T(s,t)
    + \tilde{d} w_{\Delta,1}(s,t) + \tilde{\lambda} w_{\Delta,2}(s,t),\\
&z_{\Delta,1}(s,t) = T_{ss}(s,t), 
z_{\Delta,2}(s,t) = T(s,t),z_e(t) = \int\limits_0^1 T(s,t) ds,\\ 
&y(t) = T(1,t), \quad T(0,t) = 0, \quad T_s(1,t) = 0,
\end{align*}
where \(y(t)\) is the measured output and \(z(t)\) is the regulated output. Then Monte Carlo simulations (1000 samples, Figure \ref{fig:montecarlo}) validate that the structured approach achieves $\gamma_V$ = 1.2957 versus $\gamma_V$ = 1.8002 for the unstructured case, demonstrating reduced conservatism while maintaining convergence.

\section{Conclusion}\label{sec:conc}
We have presented an integral quadratic constraint framework for robust stability and performance analysis of infinite-dimensional systems described by Partial Integral Equations. By extending the finite-dimensional IQC methodology to the PIE setting, this work provides systematic tools for analyzing spatially distributed systems subject to structured uncertainties. The proposed framework enables unified treatment of uncertain infinite-dimensional systems, robust stability and performance theorems based on Linear PI Inequalities that can be efficiently solved using PIETOOLS. The computation of structured singular value bounds for quantitative robustness analysis for the first time enables the establishment of $\mu$-theory for infinite dimensional systems.



\bibliographystyle{plain}        
\bibliography{references}  
\vspace{12pt}

\end{document}